\documentclass [12pt]{article}
\usepackage{graphicx,epsfig}
\title{Magnetic phase diagrams of the generalized spin-one-half Falicov-Kimball model}
\author{Martin \v Zonda\\
Department of Theoretical Physics and Astrophysics, Faculty of
Science,\\ P. J. \v Saf\'arik University, Park Angelinum 9, 040 01 Ko\v sice,
Slovakia}
\date{}
\begin{document}
\baselineskip=20pt
\maketitle

\begin{abstract}
A combination of small-cluster exact-diagonalization calculations and 
a well-con\-tro\-lled approximative method is used to examine the ground-state 
phase diagrams of the spin-one-half Falicov-Kimball model extended by the 
spin-dependent on-site interaction ($J$) between localized spins and itinerant
($d$) electrons, as well as by external magnetic field ($h$). Both the magnetic 
ordering and metamagnetic transitions are analysed as functions of $h$ and the number of 
itinerant electrons ($N_d$) at selected $J$.
Various magnetic superstructures including axial and diagonal spin stripes 
are observed for nonzero values of $J$ and $h$.
Moreover, it is shown that increasing $h$ strongly stabilizes 
the fully and partially polarized states, while the non polarized 
state is reduced. 
\end{abstract}

\newpage
\section{Introduction}
In the past decade, a considerable amount of effort has been devoted to 
understand the underlying physics that leads to a charge ordering in strongly 
correlated electron systems. The motivation was clearly due to the observation
of a such ordering in doped nickelate~\cite{Ni} and cuprate~\cite{Cu} materials, 
some of which constitute materials that exhibit high-temperature 
superconductivity. One of the simplest models suitable to describe charge 
ordered phases in interacting electron systems is the Falicov-Kimball model 
(FKM)~\cite{FKM}. Indeed, it was shown that already the simplest version of 
this model (the spinless FKM) exhibits an extremely rich spectrum of charge 
ordered solutions, including various types of periodic, phase-separated and 
striped phases~\cite{Lemanski,Cenci}. However, the spinless version of the FKM, 
although non-trivial, is not able to account for all aspects of real experiments. 
For example, many experiments show that a charge superstructure is accompanied
by a magnetic superstructure~\cite{Ni,Cu}. In order to describe both types 
of ordering in the unified picture, it was proposed~\cite{Results} a simple 
model based 
on a generalization of the spin-one-half FKM with an anisotropic, 
spin-dependent local interaction that couples the localized and itinerant 
subsystems. The first systematic studies~\cite{Results,Farky} clearly 
demonstrated strong cooperative effects of spin-dependent interaction on the 
ground states of the model and revealed various 
types of homogeneous as well as inhomogeneous ordering. It is naturally 
to expect,
that the formation of new structures is accompanied by changes in 
magnetization, and therefore we would like to study the ground-state 
properties
of the model in the external magnetic field. The motivation for our study
was also the work by Lema\'nski~\cite{Lem2},  where similar studies were performed 
using the restricted phase diagram method for 12 periodic configurations of 
the spins. Here we study the model exactly (over the full set of spin 
configurations) for clusters up to 36 lattice sites and approximately for 
larger clusters.\\
Our starting Hamiltonian has the form
\begin{eqnarray}
H=\sum_{ij\sigma} t_{ij}d^+_{i\sigma}d_{j\sigma} 
&+& J\sum_{i}(d^+_{i \uparrow} d_{i \uparrow} 
- d^+_{i \downarrow} d_{i \downarrow} )S^z_i\nonumber\\
&-& h\sum_{i}(d^+_{i\uparrow}d_{i\uparrow} - d^+_{i\downarrow}d_{i\downarrow})
- h\sum_{i}S^z_i,
\label{eq1}
\end{eqnarray}
where $S^z_i$ is a spin projection in $z$ direction with values $S^z_i=\pm1$,
and $d^+_{i\sigma}$,
$d_{i\sigma}$ are the creation and annihilation operators of the
itinerant electrons in the $d$-band Wannier state at site $i$.\\
\hspace*{0.5cm}
The first term of (1) is the kinetic energy corresponding to
quantum-mechanical hopping of the itinerant $d$-electrons
between sites $i$ and $j$. These intersite hopping
transitions are described by the matrix  elements $t_{ij}$,
which are $-t$ if $i$ and $j$ are the nearest neighbors and
zero otherwise (in the following all parameters are measured
in units of $t$). The second term is 
anisotropic, spin-dependent local interaction of the Ising type 
between the $d$-electrons and spins.
The last two terms represent energies of 
$d$-electrons and spins in an external magnetic field.\\
\hspace*{0.5cm}
Using the fact, that $S^z_i$ takes only two values $S^z_i=\pm1$
the  Hamiltonian (1) can be rewritten as
\begin{equation}
H=\sum_{ij\sigma} t^{\sigma}_{ij}d^+_{i\sigma}d_{j\sigma}
- h\sum_{i}S^z_i,\\
\label{eq2}
\end{equation}
where $t^{\sigma}_{ij}=t_{ij}+\sigma(JS^z_i-h)\delta_{ij}$. 
Thus for a given spin distribution $s=\{S^z_1,S^z_2,...,S^z_L\}$,
the Hamiltonian (2) is the second-quantized version 
of the single particle Hamiltonian $t^{\sigma}(s)$, so the investigation of 
the model (2) is reduced to the investigation of the spectrum of $t^{\sigma}$ 
for different configurations of spins. This can be done exactly over the full 
set of spin configurations or approximately over a reduced set. In this paper 
we present a combination of both methods. For clusters up to $L=36$ lattice 
sites we use small-cluster exact-diagonalization calculations and 
for larger clusters we use the well-controlled numerical method 
described in detail in the papers~\cite{method}.

\section{Results and discussion}
In the present work we study the one ($D=1$) and two ($D=2$) dimensional 
analogue of the model for the spin interaction value $J=0.5$ and for a wide 
range 
of magnetic field $h$ (from $h=0$ to $h=0.5$ with step 0.01). To reveal the 
finite-size effects numerical calculations were done on different clusters. 
We have found, that the main features of the phase 
diagrams in the $N_d-h$ plane (for $D=1$ as well as $D=2$) hold on all 
examined lattices and thus can be used
satisfactorily to represent the behaviour of macroscopic systems. 
Here we present the one dimensional results for $L=24$, that represent the 
typical one-dimensional behaviour, while for $D=2$ the numerical results are 
presented for $L=36$ (the largest cluster that we were able to consider exactly).\\
Let us start a discussion of the phase diagram with 
a description of configuration types that form its basic structure in one 
dimension. Manifold phases entered into the phase diagrams are classified 
according to $S^z=\sum_iS^{z}_{i\uparrow}-S^{z}_{i\downarrow}$ and
$S^z_d=N_{d\uparrow}-N_{d\downarrow}$: the fully polarized (FP) phase 
characterized by $|S^z|=L$, $|S^z_d|=N_d$, the partially polarized (PP) phases
characterized by $0<|S^z|<L$, $0<|S^z_d|<N_d$ and the non polarized (NP) phases
characterized by $|S^z|=0$, $|S^z_d|=0$. 
Comparing these phase diagrams with ones obtained for $h=0$~\cite{Farky}, one 
can see rather different behaviours. Indeed, while for $h=0$ the magnetic 
phase diagrams coincide practically over the whole range of model parameters, 
the non-zero $h$ destroys obviously this coincidence. Now the phase diagrams 
coincide only at low $d$-electron concentrations where the ground state is 
the FP phase for both the $d$-electron as well as spin subsystems. But with 
increasing $N_d$ the spin and $d$-electron systems are developed differently. For 
higher values of $N_d$ the spin subsystem still prefers the FP state, contrary 
to the $d$-electron subsystem where the PP state is stabilized. A different 
behaviour is observed also for $h\rightarrow$ 0, where the spin subsystem prefers 
PP state, while the $d$-electron subsystem prefers the NP state. Besides these 
differences the phase diagrams of spin and $d$-electron subsystems exhibit one 
same characteristic and namely that very small changes of $h$ can produce
large cooperative changes, indicating possible metamagnetic transitions. 
Before discussing  these transitions let us discuss the complete set of
ground state configurations, that enter to the spin phase diagram.
The complete set of ground-state configurations consists of only a few types,
including 7 different NP phases, 10 different PP phases and of course the FP 
phase.
Between them one can find different types of periodic or perturbed periodic 
and non-periodic 
configurations, but the most interesting examples represent NP phases. In this
case all configurations (with the exception of $N_d=22$) have a central point 
symmetry.

Let us now turn on our attention to the problem of metamagnetic transitions. To 
reveal the structure of magnetization curves we have performed exhaustive 
studies of the model for two different $d$-electron concentrations
$n_d=1/2$ and $n_d=3/4$. In Fig.~2 we summarize numerical results for both
electron as well as spin subsystems. In the first case ($n_d=1/2$) the spin 
magnetization curve consists of two significant stairs 
(see Fig.~2a) $m_f=0$ and $m_f=1$ 
(or  $m_d=0$ and $m_d\sim 0.5$ for $d$-electrons, see Fig.~2b). It is very 
interesting, that 
the relative small changes  of $h$ lead to the significant spin 
reorientation from the periodic arrangement 
$\{\uparrow\uparrow\downarrow\downarrow
\uparrow\uparrow......\downarrow\downarrow\uparrow\uparrow\downarrow
\downarrow\}$ ($m_f=0$) to the FP state ($m_f=1$). 
The similar situation holds also for $n_d=3/4$, 
where the transition from the NP phase to the FP phase realizes through the PP states
(Fig.~2c). In the spin magnetization curve we have observed four significant 
stairs ($m_f=$ 0, 1/4, 1/3 and 1) corresponding to four different configuration 
types 
($w_0=[\uparrow\uparrow\downarrow\uparrow\downarrow\downarrow\uparrow\downarrow]_{L/8},
w_{1/4}=[\uparrow\uparrow\downarrow\uparrow\uparrow\downarrow\uparrow\downarrow]_{L/8},
w_{1/3}=[\uparrow\uparrow\downarrow]_{L/3}$
and $w_1=\{\uparrow...\uparrow\}$, where the lower index denotes the number of repetitions
of the block [...]). The same structure is observed also on the $d$-electron 
magnetization curve as illustrates Fig.~2d.

Let us now briefly discuss  the case $D=2$. In Fig.~3 we present the phase 
diagrams of the model in the $N_d-h$ plane obtained by exact-diagonalization 
calculations for $L=36$ and $J=0.5$. Comparing these phase diagrams with their 
one-dimensional counterparts one can find obvious similarities. Again the phase 
diagrams coincide only at very low $d$-electron concentrations 
where the ground state is the FP phase, while for increasing $N_d$
the spin ($d$-electron) subsystem prefers the FP (PP) state.
The NP ground states for non-zero $h$ are observed only on isolated lines (points)
at 
$N_d$: $N_d=18, 26$ for spins and $N_d$= 2, 10, 18, 26 and 36 for $d$-electrons.\\
The two-dimensional results are of particular importance since they could shed
light on the mechanism of two-dimensional charge and magnetic ordering in 
doped nickelate~\cite{Ni} and cuprate~\cite{Cu} materials. Here we concern our 
attention on a description of basic types of magnetic ordering 
that exhibits the spin-one-half FKM with spin-dependent interaction between 
$d$-electrons and spins in the magnetic field $h$ for $D=2$.
Analysing obtained numerical data, it was found  
that the set of ground-state configurations consists of only
a few basic configuration types, listed in Fig.~4.\\ 
With increasing $N_d$
one can see a general tendency of the system to change the ground state from the 
phase  separation (observed for small $N_d$) through special inhomogeneous 
axial stripes (intermediate $N_d$) to the regular "chessboard" structure 
(around $N_d=L$). Moreover, the unusual central point symmetry  observed 
for NP phases in $D=1$ case persists also for $D=2$ (for all NP states 
with the exception of $N_d=10$ and $N_d=22$), of course now as a mirror symmetry.
Again, we have studied the magnetizations for spin as well as $d$-electron 
subsystems on clusters up to $L=100$ lattice sites. Unfortunately, we were 
not able to obtain the definite results, due to finite-site effects. 

In summary, a combination of small-cluster exact diagonalization calculations 
and a well-controlled approximative method was used to study the ground-state
phase diagrams of the generalized spin-one-half FKM with spin-dependent on-site
interaction $J=0.5$ extended by magnetic field for one and two dimensional cases.
For both cases it was found that the magnetic field stabilizes the FP 
(for spins) or PP/FP states (for $d$-electrons), while the NP states persist 
only for $h\rightarrow 0$. Although, the non-zero $h$ destroys the coincidence 
between the spin and $d$-electron subsystems (observed for $h=0$), the one common
feature is evident, and namely, that the very small changes of $h$ can produce
large cooperative changes in the ground state of the model. This observation
was supported also by studies of magnetization curves, where we have observed strong
changes from the periodic NP/PP state to the FP state 
as well as with configuration type analysis, where we have identified 
phase separated, striped, periodic and non-periodic spin distributions.

\newpage

\begin{figure}[h]
\begin{center}
\includegraphics[width=16cm]{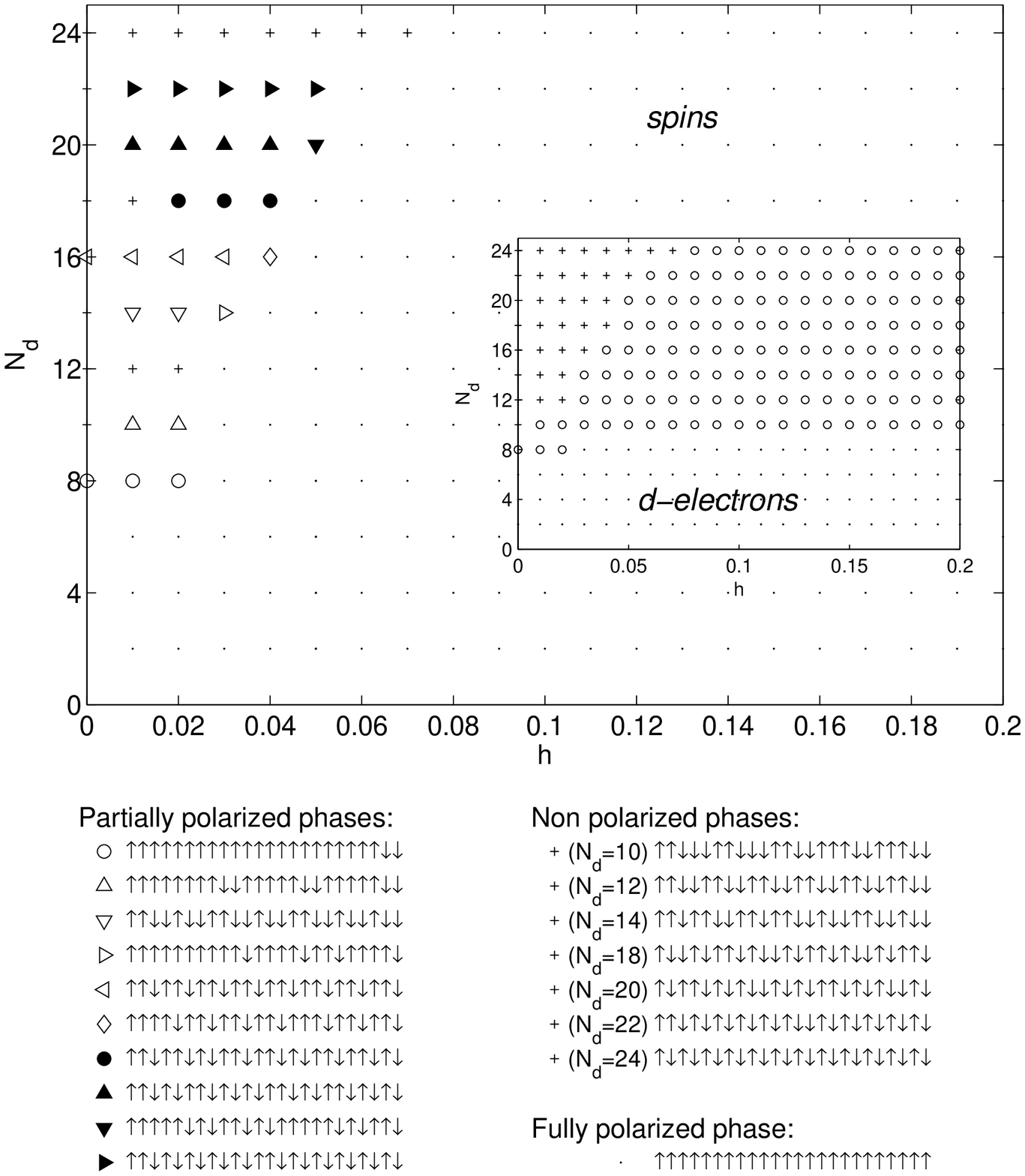}
\end{center}
\caption{Spin phase diagram of the model for $J=0.5$ in the $N_d-h$ plane for 
$L=24$ and $D=1$. 
Inset: Magnetic phase diagram of the $d$-electron subsystem. Different symbols 
represent different magnetic phases: FP (.), PP ($\circ$) and NP (+) phases.}
\label{fig:1}
\end{figure}

\begin{figure}[h]
\begin{center}
\includegraphics[width=15cm]{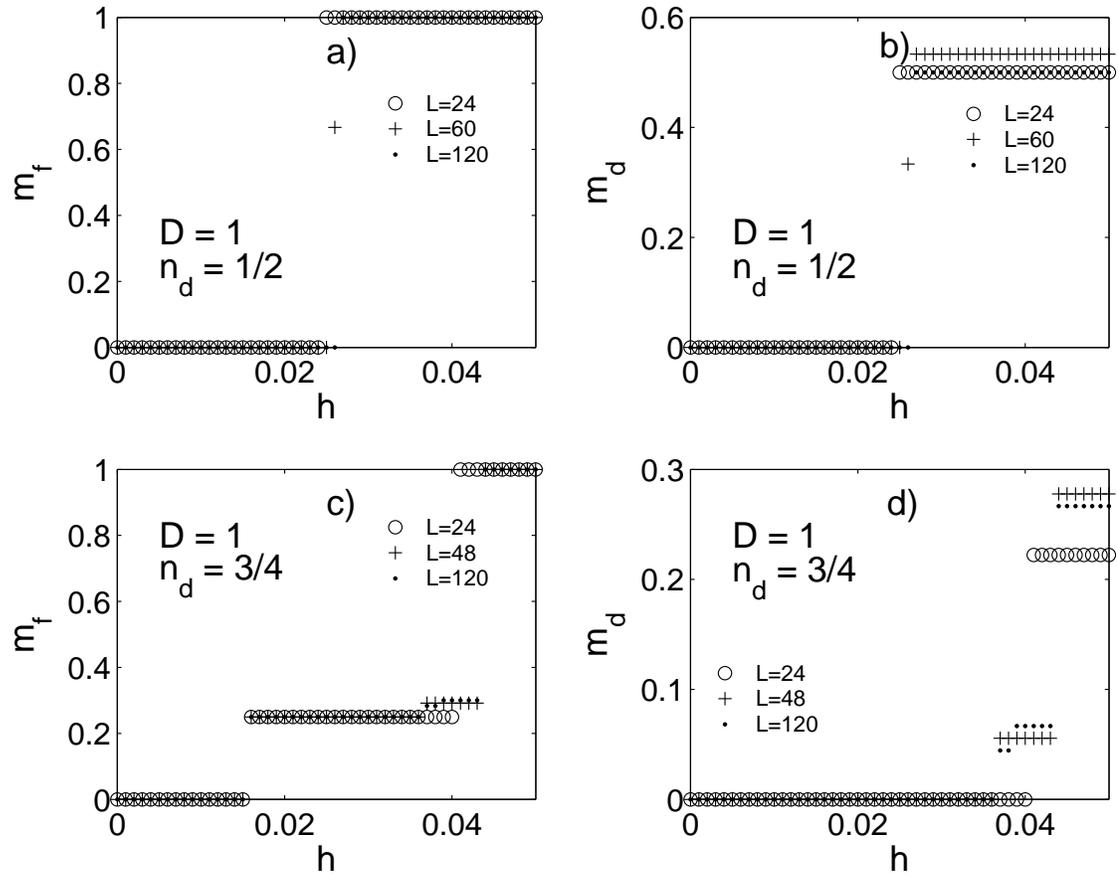}
\end{center}
\caption{a) The total magnetization of the spins for $J=0.5$, $n_d=1/2$ and
different clusters ($L=24, 60$ and $120$) in $D=1$. b) Magnetization of 
electrons for the same parameters. c) The total magnetization of the spins for 
$J=0.5$, $n_d=3/4$ and different clusters ($L=24,48$ and $120$) in $D=1$. 
d) Magnetization of electrons for the same parameters. }
\label{fig:2}
\end{figure}

\begin{figure}[h]
\begin{center}
\includegraphics[width=15cm]{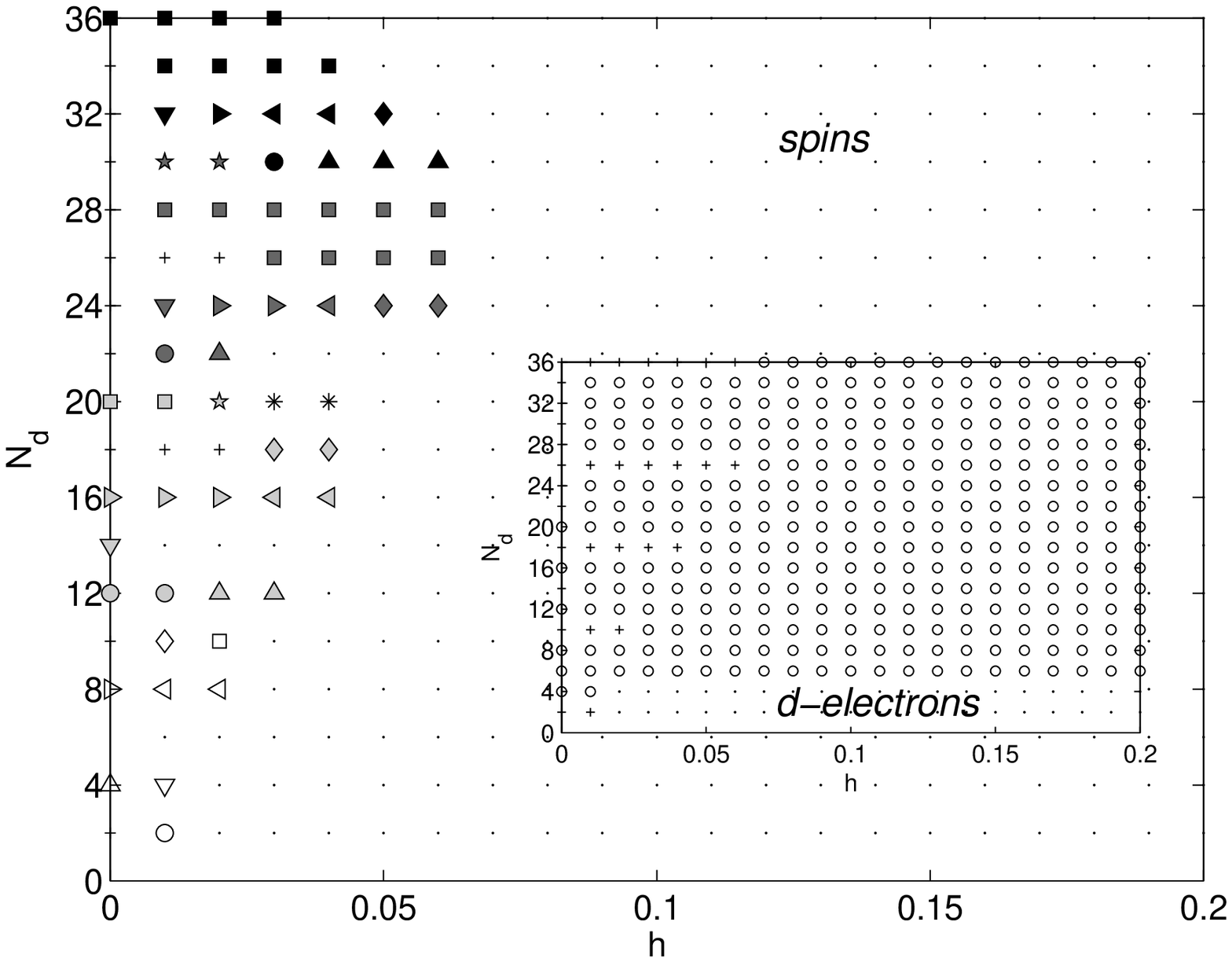}
\end{center}
\caption{Spin phase diagram of the model for $J=0.5$ in $N_d-h$ plane for 
$L=36$ and $D=2$. All ground-state configurations (with the exception of 
FP (.)) are 
displayed in Fig.~4.
Inset: Magnetic phase diagram of the $d$-electron subsystem. Different symbols 
represent different magnetic phases: FP (.), PP ($\circ$) and NP (+) phases.}
\label{fig:3}
\end{figure}

\begin{figure}[h]
\begin{center}
\includegraphics[width=13cm]{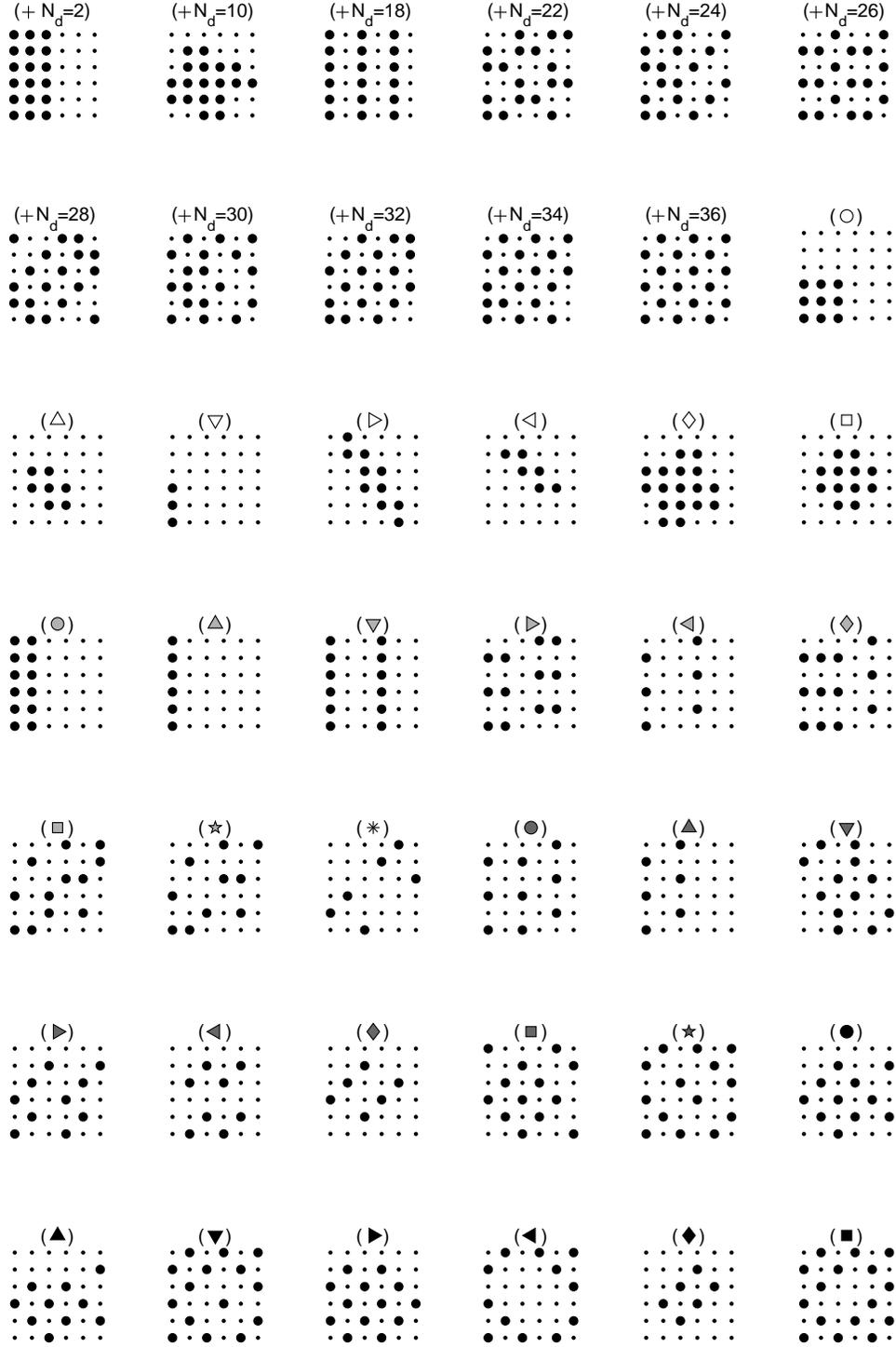}
\end{center}
\caption{Arrangements of localized spins that enter into the phase diagram 
displayed in Fig.~3. First 11 phases correspond with NP phases. All other
phases are PP. To visualize the spin distributions we use ($\bullet$) for the 
up spin orientation and (.) for the down spin orientation.}
\label{fig:4}
\end{figure}

\end{document}